\documentclass{article} % One-column default

\usepackage{preprint}
\definecolor{cardinal} {RGB}{196, 30, 58}
\definecolor{lightgrey}{RGB}{150,150,150}

\usepackage[role = cardinal, grid = lightgrey]{credits}
\usepackage{listings, xcolor}

\definecolor{verylightgray}{rgb}{.97,.97,.97}

\lstdefinelanguage{Solidity}{
	keywords=[1]{anonymous, assembly, assert, balance, break, call, callcode, case, catch, class, constant, continue, constructor, contract, debugger, default, delegatecall, delete, do, else, emit, event, experimental, export, external, false, finally, for, function, gas, if, implements, import, in, indexed, instanceof, interface, internal, is, length, library, log0, log1, log2, log3, log4, memory, modifier, new, payable, pragma, private, protected, public, pure, push, require, return, returns, revert, selfdestruct, send, solidity, storage, struct, suicide, super, switch, then, this, throw, transfer, true, try, typeof, using, value, view, while, with, addmod, ecrecover, keccak256, mulmod, ripemd160, sha256, sha3}, % generic keywords including crypto operations
	keywordstyle=[1]\color{blue}\bfseries,
	keywords=[2]{address, bool, byte, bytes, bytes1, bytes2, bytes3, bytes4, bytes5, bytes6, bytes7, bytes8, bytes9, bytes10, bytes11, bytes12, bytes13, bytes14, bytes15, bytes16, bytes17, bytes18, bytes19, bytes20, bytes21, bytes22, bytes23, bytes24, bytes25, bytes26, bytes27, bytes28, bytes29, bytes30, bytes31, bytes32, enum, int, int8, int16, int24, int32, int40, int48, int56, int64, int72, int80, int88, int96, int104, int112, int120, int128, int136, int144, int152, int160, int168, int176, int184, int192, int200, int208, int216, int224, int232, int240, int248, int256, mapping, string, uint, uint8, uint16, uint24, uint32, uint40, uint48, uint56, uint64, uint72, uint80, uint88, uint96, uint104, uint112, uint120, uint128, uint136, uint144, uint152, uint160, uint168, uint176, uint184, uint192, uint200, uint208, uint216, uint224, uint232, uint240, uint248, uint256, var, void, ether, finney, szabo, wei, days, hours, minutes, seconds, weeks, years},	% types; money and time units
	keywordstyle=[2]\color{teal}\bfseries,
	keywords=[3]{block, blockhash, coinbase, difficulty, gaslimit, number, timestamp, msg, data, gas, sender, sig, value, now, tx, gasprice, origin},	% environment variables
	keywordstyle=[3]\color{violet}\bfseries,
	identifierstyle=\color{black},
	sensitive=true,
	comment=[l]{//},
	morecomment=[s]{/*}{*/},
	commentstyle=\color{gray}\ttfamily,
	stringstyle=\color{red}\ttfamily,
	morestring=[b]',
	morestring=[b]"
}

\usepackage{float}
\usepackage{adjustbox}
\usepackage{array}
\usepackage{booktabs}
\usepackage{multirow}
\usepackage{graphicx}
\usepackage{color}
\usepackage{soul}
\usepackage{microtype}
\usepackage{paralist}
\usepackage{cite}
\usepackage{amsmath,amssymb,amsfonts}
\usepackage{algorithmic}
\usepackage{textcomp}
\usepackage{xcolor}
\usepackage{pgf-pie}
\usepackage{tikz}
\usepackage{hyperref}
\usepackage{url}
\def\BibTeX{{\rm B\kern-.05em{\sc i\kern-.025em b}\kern-.08em
    T\kern-.1667em\lower.7ex\hbox{E}\kern-.125emX}}

\definecolor{lightgrey}{rgb}{0.925, 0.925, 0.925}
\sethlcolor{lightgrey}
\makeatletter
\def\SOUL@hlpreamble{%
    \setul{}{3.5ex}% increased by 1ex
    \let\SOUL@stcolor\SOUL@hlcolor
    \dimen@\SOUL@ulthickness
    \dimen@i=-.75ex % increased by -0.25ex
    \advance\dimen@i-.5\dimen@
    \edef\SOUL@uldepth{\the\dimen@i}%
    \let\SOUL@ulcolor\SOUL@stcolor
    \SOUL@ulpreamble
}
\makeatother

\newcommand{\para}[1]{\smallskip\noindent\textbf{#1}}
\newcommand{\tuint}{\texttt{uint}}
\newcommand{\tbyte}{\texttt{byte}}
\newcommand{\tstring}{\texttt{string}}

\credit{R.K}{1,1,1,1,0,0,1,1,1,1}
\credit{E.L}{0,0.7,0,0,0,0,0,0.8,0,0}
\credit{L.X and }{0,0,0,0,0,0.65,1,0,0,1}
\credit{W.S}{0,0,0,0,1,1,1,0,0,0.6}

\title{Who is Smarter? An Empirical Study of AI-based Smart Contract Creation}

\newif\ifuniqueAffiliation
\uniqueAffiliationtrue

\ifuniqueAffiliation % Standard variant of author block
\author{ \href{https://orcid.org/0000-0002-6705-6506}{\includegraphics[scale=0.06]{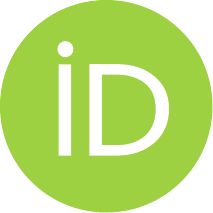}\hspace{1mm}Rabimba Karanjai}\thanks{www.rabimba.me} \\
	Department of Computer Science\\
	University Of Houston\\
	\texttt{rkaranjai@uh.edu} \\
	\And
	Edward Li \\
	\texttt{edward.w.li@outlook.com} \\
        \And
	Lei Xu \\
 	Department of Computer Science\\
        Kent State University \\
	\texttt{xuleimath@gmail.com}\\
        \And
	Weidong Shi \\
 	Department of Computer Science\\
        University Of Houston \\
	\texttt{wshi3@uh.edu}\\
}
\else
\usepackage{authblk}

\newbox{\orcid}\sbox{\orcid}{\includegraphics[scale=0.06]{orcid.pdf}} 
\author[1]{%
	\href{https://orcid.org/0000-0000-0000-0000}{\usebox{\orcid}\hspace{1mm}David S.~Hippocampus\thanks{\texttt{hippo@cs.cranberry-lemon.edu}}}%
}
\author[1,2]{%
	\href{https://orcid.org/0000-0000-0000-0000}{\usebox{\orcid}\hspace{1mm}Elias D.~Striatum\thanks{\texttt{stariate@ee.mount-sheikh.edu}}}%
}
\affil[1]{Department of Computer Science, Cranberry-Lemon University, Pittsburgh, PA 15213}
\affil[2]{Department of Electrical Engineering, Mount-Sheikh University, Santa Narimana, Levand}
\fi

\begin{document}
\maketitle  
% \thispagestyle{plain}
% \pagestyle{plain}
% typeset the header of the contribution
%

\begin{abstract}
The advent of advanced language models, such as ChatGPT and Google Palm2, for crafting smart contracts marks a pioneering step towards AI-driven pair programming. These models boast an extensive repository of open-source smart contracts, giving them an edge in generating Solidity code over other tools.

While preliminary evaluations of these models in smart contract creation show potential, a comprehensive analysis is essential to truly gauge their strengths and limitations. This research seeks to scrutinize the caliber of smart contract code produced by these models. Additionally, we delve into understanding how the quality and diversity of input parameters influence the output. For this purpose, we've designed a rigorous testing framework to assess the validity, accuracy, and efficiency of the generated code.

Our findings highlight significant security vulnerabilities in the produced smart contracts and raise concerns about the overall code quality and accuracy. Nevertheless, we've pinpointed avenues for enhancement. This document also outlines prospective research pathways to refine the methodology, quality, and security of the smart contract codes produced.
\end{abstract}

\section{Introduction}
%Large language models are a significant breakthrough in deep learning, as they can be used for a variety of tasks such as text translation~\cite{chung2022scaling,brown2020language} and summarization, as well as chatbots like Chat-GPT. As these models grow in size, they become more capable of learning and can even be applied to tasks involving code generation and understanding.

In recent years, large language models (LLMs) such as generative pre-trained transformer (GPT) models~\cite{brown2020language} and Google PaLM2~\cite{anil2023palm} are becoming more mature and ready for deployment for real applications~\cite{https://doi.org/10.48550/arxiv.2302.03202,brown2020language}.
Among all the applications enabled by LLMs, machine-aided source code generation from natural language is fascinating and widely believed to have the potential to revolutionize the way humans create programs.
For instance, ChatGPT and its derivatives (e.g., Github CoPilot~\cite{10.1145/3511861.3511863}) have been used to generate source code of multiple programming languages based on natural language inputs from the programmer. But there have also been concerns about these being used for malicious purposes \cite{karanjai2022targeted}.

One of the major concerns of LLM-generated source code is its safety because there is a lack of rigorous understanding of the logic that the source code is generated.
Many works have been done on analyzing the security features of the generated code for languages such as Python, Golang, and Javascript.
In most cases, LLMs rely on the training dataset to ``understand'' natural language instructions to create source code and are programming language specific, i.e., they can only create source code of programming languages that have been used to train the model. 
Accordingly, the security analysis works are also language specific.
And most security assessment results on one programming language may not be valid for another language.

Recently, decentralized ledgers and smart contracts running atop it are finding more applications, especially in the financial sector. 
For instance, the Uniswap smart contracts process about \$7.17 billion per day in 2021~\cite{uniswap}.
Because of the popularity of smart contracts and their importance in various scenarios like Confidential Computing \cite{10.1145/3505253.3505259,10174906}, Decentralized Serverless Functions~\cite{10174945}, Event based Transactions \cite{10.1145/3464298.3493401,9461133}, it is a natural question to ask whether LLMs can do a good job at creating smart contracts based on user instructions and how secure the generated smart contracts are.
%
%The existing work has mostly focused on the security assessment of these generative models for code generation targeting some specific programming languages like Python, Golang, and JavaScript. There is a need to broaden the study of the generative models for code generation to cover more diverse programming languages as well as assess the generated software from more comprehensive dimensions beyond security risk. 
%
To fill the gap, we investigate two popular LLMs, ChatGPT and Google PaLM2~\cite{anil2023palm}, for generating smart contract code in Solidity, which is a mainstream smart contract programming language.  
We compare the code generation process and the generated smart contracts, and conduct systematic evaluation on the contracts. 
%add some details of the results
To systematically evaluate GPT 3.5 and PaLM2, we propose to construct an experimental setup to assess the generated code in terms of validity, correctness, and efficiency. In this context, we defined the following research questions:
\begin{itemize}
    \item \textbf{RQ1} What is the quality of the code generated by LLMs like ChatGPT and Google Palm2?
    \begin{itemize}
        \item \textbf{RQ1.1} Code Validity
        \item \textbf{RQ1.2} Code Correctness
    \end{itemize}
    \item \textbf{RQ2} Can better code be generated using different prompting techniques
    \item \textbf{RQ3} Are the generated codes secure?
\end{itemize}

In summary, the main contributions of the paper are:
\begin{inparaenum}[(i)]
    \item We have explored and evaluated the feasibility and quality of the models for creating Solidity smart contract code;
    \item We tested and compared two popular source code generation tools, ChatGPT and Google PALM2 for smart contract code generation;
    \item We systematically assessed and evaluated the contract code generated by the GPT models and discussed insights obtained from the experiments; and
    \item In addition, we highlight possible directions for further research. 
\end{inparaenum}

%Also in line with our studies, we try to answer the following research questions:

% \section{Related Work}\label{sec-related}
%Although these works are impressive and have been proven useful in practice, little eff

\section{Background}\label{sec-background}

Assessing the quality of the software and identifying bugs as well as security vulnerabilities before deployment can prevent potential harm and unexpected expenses. Nevertheless, automatically detecting security-critical bugs in code is a difficult task in reality. This also applies to code that is generated by the AI models, particularly due to the complex and opaque nature of such models (treated as a black box). When it comes to Solidity, it is more important to investigate both code correctness, any kind of bugs, and performance-related issues like bytecode size and gas fee cost that any AI-based automated systems can create.

\subsection{Ethereum, Smart Contract and Its Security}
Ethereum is a widely used public, decentralized ledger that keeps immutable and transparent records that can be programmed on the ledger. These programmable records are known as smart contracts. Smart contracts enable the sharing and execution of program logic among multiple parties. They are typically written in Solidity, which is a programming language specifically designed for smart contracts. Solidity follows an object-oriented approach, uses static typing, and compiles to bytecode that can be executed on Ethereum Virtual Machine (EVM).
%Solidity is a programming language used to write smart contracts. It is a statically typed, object-oriented language with support for inheritance, libraries, and other features. 
%In line 2, a contract called "Register" is defined, which allows a student to register their name. Before that, in line 1, the Solidity compiler version is specified. 
As Solidity compilers change rapidly, the latest version may not be backward-compatible with all smart contracts. In this case, the acceptable compiler versions are between 0.8.6 and 0.9.0.

The Solidity language supports various built-in types, including \tuint, \tbyte, and \tstring. It also has special types such as addresses. The address is a unique global identifier for all EOAs (Externally Owned Accounts) and smart contracts. All Ether transfers and contract executions must target a specific address. Solidity also has a built-in key-value dictionary called ''mapping''. 

% Solidity has access modifiers to control accessibility to state variables and functions. The public modifier allows both internal and external access to variables and functions. The private modifier only prevents access from other addresses inside the blockchain but cannot prevent users from reading the value from outside the blockchain. The header of a function is an external function that can only be called from outside the contract or through this (a reference to the current contract instance). %The header also indicates that the function returns a Boolean value.

%\subsubsection{Bugs and Security Vulnerabilities in Solidity}
According to the security analysis of Ethereum~\cite{kim2020automated}, the majority of vulnerabilities on the platform are found in the application layer, particularly in smart contracts where they should be avoided. Smart contract vulnerabilities can range from common issues such as integer overflow to more specific problems related to the blockchain, such as re-entrancy. Despite being challenging to exploit, blockchain-specific vulnerabilities have resulted in severe attacks such as the well-known TheDAO~\cite{thedao} attack, largely because they were previously unknown. 
%In this section, we will provide a brief overview of some of the typical vulnerabilities that are commonly found in smart contracts.

\subsection{Tools for Evaluating Security Issues}
Several security testing techniques can be utilized to effectively discover software vulnerabilities and avoid bugs during the operation of a deployed system \cite{beller2016analyzing,chatzieleftheriou2011test,christakis2016developers}. 
To accomplish this objective, these techniques aim to identify various programming errors, inadequate coding style, deprecated functionalities, or possible memory safety violations (such as unauthorized access to unsafe memory that can be exploited after deployment or outdated cryptographic schemes that are shown to be insecure~\cite{gosain2015static,goseva2015capability,szekeres2013sok}). 
Broadly, current methods for assessing the security of software can be divided into two categories: static~\cite{beller2016analyzing,ayewah2008using} and dynamic analysis~\cite{fioraldi2020afl++,shoshitaishvili2016sok}. 
Static analysis examines the code of a given program to detect potential vulnerabilities, while the latter approach executes the code. For instance, fuzz testing (fuzzing) produces random program executions to trigger bugs in the program.

For our work, we have taken the approach of using static analysis tools to check for bugs in the GPT model-generated contract code. The detected bugs have been categorized into Low, Medium, and High severity categories based on their security implications of them. We have used a mix of open-source and commercial static analysis tools as described in Section~\ref{tool}.

% \subsection{Code Generation}
% Large language models represent a significant breakthrough in contemporary deep learning advancements. As they become larger in size, their learning capacity enables them to be used for various tasks, including generating code for AI-assisted pair programming. Given a prompt that describes the desired function, the model can generate appropriate code. For our work we concentrate on Codex \cite{chen2021evaluating}.

\subsection{Large Language Models and Prompts}
The natural language processing field has been advanced by LLMs, which have demonstrated significant progress in tasks such as question answering, translation, and reading comprehension~\cite{brown2020language,raffel2020exploring}. 
This progress has been made possible by increasing the model size from hundreds of millions~\cite{kenton2019bert} to hundreds of billions~\cite{brown2020language}, using self-supervised objective functions, and utilizing large corpora of text training data. 
These models are often developed by large corporations and made available as pre-trained models. Brown et al.~\cite{brown2020language} have demonstrated that these models can be applied to various tasks with just a few input examples, without needing to adjust the model parameters. Users can provide a few-shot prompt template to guide the models toward generating the desired output for a particular task. 

\subsection{LLM and Code Generation}
There is a growing interest in using large language models to understand and generate source code \cite{chen2021evaluating,fried2022incoder,wang2021codet5}. Feng et al.~\cite{feng2020codebert} and Guo et al. \cite{guo2020graphcodebert} have proposed encoder-only models with a modified objective function, focusing on code classification, code retrieval, and program repair \cite{feng2020codebert,guo2020graphcodebert}. Ahmad et al. \cite{ahmad2021unified} and Wang et al. \cite{wang2021codet5} have used an encoder-decoder architecture to address code-to-code and code-to-text generation tasks, such as program translation, program repair, and code summarization. Recently, decoder-only models have shown promise in generating programs in a left-to-right sequence \cite{chen2021evaluating,nijkamp2022codegen}. These models can be used for zero-shot and few-shots program generation tasks, such as code completion, code infilling, and text-to-code conversion \cite{chen2021evaluating,fried2022incoder,nijkamp2022codegen}. 

% \textit{Codex:} The Codex model is based on GPT-3 \cite{brown2020language}, which is a transformer-based autoregressive language model trained on natural text. To fine-tune Codex, 54 million public software repositories available on GitHub were utilized in the training set. The resulting Codex model has 12 billion parameters and is utilized for GitHub CoPilot.

\section{Code Generation in Smart Contract}

% \subsection{Prompt Engineering}
Prompt engineering for creating issue-free Solidity smart contracts poses several challenges. One of the main challenges is the need for a deep understanding of the Solidity programming model and its associated issues including security risks. Solidity has its unique features and syntax, making it different from other programming languages, and thus requires specific expertise and insight to develop high-quality code. Additionally, Solidity smart contracts can suffer from various vulnerabilities, such as integer overflow/underflow, re-entrancy, and logic errors, among others. Another challenge is the difficulty in defining clear and concise prompts that accurately capture the intended functionality of the smart contract. The lack of clarity in prompts may lead to misunderstandings, which can result in incorrect implementation, inefficiency, and subsequent vulnerabilities. Addressing these challenges will require expertise in both Solidity programming and prompt engineering to ensure that the resulting smart contracts are issue-free, acceptable quality, and secure.

For the scope of the paper, we take a guided approach to generating prompts which give us a better performance as described in Section~\ref{prompt-eng}.

% use two approaches, which are described below:
% \begin{itemize}
%     \item \textit{Full text prompt.}  Since ChatGPT (V2) supports full-sentence explanatory prompts, these were used for generating smart contracts. Most of the time careful prompt formulation was needed to generate compilable and functional code. This has been detailed more in Section~\ref{prompt-eng}.
%     \item \textit{Partial comments.} Since GitHub CoPilot does not support writing full sentences as a proper way to prompt in generating compilable code. We resorted to individual smaller snippets of prompts to generate the same equivalent prompt as our ChatGPT test case.
% \end{itemize}

\begin{table}
\centering
\caption{Comparing Relevant Code Generation Tools}
\label{tab:comparison-table}
%\scriptsize
\footnotesize
\resizebox{\linewidth}{!}{%
\begin{tabular}{|>{\hspace{0pt}}m{0.388\linewidth}||>{\hspace{0pt}}m{0.313\linewidth}|>{\hspace{0pt}}m{0.237\linewidth}|} 
\hline
\multicolumn{1}{|>{\hspace{0pt}}m{0.388\linewidth}|}{Features} & ChatGPT & Google Plam2 \\ 
\hline
IDE Support & No IDE Support & No IDE Support \\ 
\hline
First Release Time & Nov 30,2022 & May 10, 2023 \\ 
\hline
Developer & OpenAI & Google \\ 
\hline
Providing References to Suggestions & NO & YES \\ 
\hline
Explanation of Suggestions & YES & YES \\ 
\hline
Providing Multiple Suggestions & NO~ & YES~ \\ 
\hline
Training Data Source & GitHub Repositories & “Google licensed code" \\ 
\hline
Multipurpose (other than programming) & YES & YES \\ 
\hline
Subscription & ChatGPT Free \par{} ChatGPT Plus (\$20 per month) & Insider Beta Access \\ 
\hline
Can be Used Offline? & NO & NO \\ 
\hline
Can it Access Local Files? & NO & YES \\
\hline
\end{tabular}
}
\end{table}

\section{Methodology}
In this section, we present our experimental setup. We commence by elaborating on the details of the experimental framework. After that, we present how we % in the models and 
assess the effectiveness and scalability of our proposed method. We also explore the transferability of the prompts generated by our approach among various models. Additionally, we demonstrate that our approach can detect security vulnerabilities in the code generated by the GPT models. 

\subsection{Dataset}

For our experiment, we could not use any of the existing datasets like HumanEval~\cite{chen2021codex} since they primarily concentrate on Python as a programming language. To evaluate we had to create our dataset. To evaluate the correctness of our code we carefully curated a set of source files from GitHub under open source permissive license. 

We leveraged GitHub APIs to construct our scraper that collects Solidity code from different projects. There were several factors we took into account when collecting the corpus:

\begin{itemize}
    \item The project has to have 50 stars on GitHub. This would ensure the project has enough interest, is a fairly popular project, and has a higher chance of having human audience code
    \item The projects should have "Solidity" as a programming tag available for further filtering
    \item They have a sufficient amount of comments present in them. This helps us design and develop prompts in scale and create the experimentation pipeline.
\end{itemize}

Once these projects were collected, they are further processed to exclude any files which were not Solidity. 
A brief overview of the files we have collected is given below.

\para{Raw code corpus collection.} 
GitHub serves as an excellent resource for accessing a wide range of publicly accessible source code written in various programming languages. In our process, we replicated the most widely recognized repositories that were labeled with the Solidity language tag and garnered at least 50 stars. From each project, we extracted all the files that were authored in Solidity, thereby generating the primary training dataset.

\para{Data pre-processing.} 
The detailed data pre-processing strategy comparison with other models is analyzed in \tablename~\ref{tab:comparison-table}.

% \begin{table}[ht]
% \centering
% \caption{Comparison of data preprocessing strategies of different models}
% \begin{tabular}{p{0.3\linewidth}p{0.3\linewidth}p{0.4\linewidth}}
% \hline
% Model & Deduplication & Filtering \\ \hline
% PolyCoder & Exact & Files >\ 1 MB, \< 100 tokens \\
% CodeParrot & Exact & Files \> 1MB, max line length \> 1000, \\
% & & mean line length \> 100, fraction of \\
% & & alphanumeric characters \< 0.25, \\
% & & containing the word "auto-generated" \\
% & & or similar in the first 5 lines \\
% Codex & Unclear, mentions “unique” & Files > 1MB, max line length > \\
% & & 1000, mean line length > 100, \\
% & & autogenerated (details unclear), \\
% & & contained a small percentage of alphanumeric characters (details unclear) \\ 
% Our Dataset & Exact & No Filtering \\ \hline
% \end{tabular}
% \end{table}

All our available codes were passed through a parser that would strip out the functions from the Solidity files, \textbf{however keeping the pragma, variables and other code information intact. Including comments.} These processed files are called \textit{Code templates} in the remainder of the paper.

\subsection{Code Analysis}

We analyze the generated codes based on our predefined parameters. To answer our different RQ's we have carefully designed an experimental workflow defined in \figurename~\ref{experimental_workflow}. 

Once the code generation part finishes, we pass it through the compilation stage and different code quality analysis steps. Once they are done we also do static analysis on them.

\subsubsection{Tools for Analyzing Smart Contracts}\label{tool}
This section introduces the tools we used for our smart contract analysis.

\textit{Mythril}~\cite{mythril}, which is a Python-based command-line tool. It conducts an interactive analysis of EVM bytecode and presents the Control Flow Graph (CFG) visually. Nodes in the CFG display disassembled code and edges are labeled by path formulas. The tool uses the Z3 SMT solver to prune the search space and compute concrete values to exploit potential vulnerabilities. Detailed documentation of the checked vulnerabilities is available online. Developed and maintained by ConsenSys, Mythril has been available on GitHub under an MIT license since September 2017.
%

% \textit{Oyente}~\cite{luu2016making} is another well-established tool for smart contract analysis, which has been used as a starting point for several other projects. It analyzes EVM bytecode symbolically and detects issues such as Ether flow influenced by transaction order, computations dependent on block timestamps, uncaught exceptions raised by calls, or potential contract re-entrancy. Oyente uses the SMT solver Z3 to prune unreachable parts of the search space. To obtain the bytecode, it relies on the Solidity compiler, and for displaying symbolic opcodes, it uses the disassembler from Go-Ethereum. The tool is written in Python and available on GitHub under a GPL-3.0 license since January 2016. Although primarily a command-line tool, it also provides a web interface.
% %

\textit{Slither}~\cite{kim2020automated} is a tool designed for the static analysis of Solidity source code that is version 0.4 or higher. It offers APIs for users to customize their analysis of the source code and generates its own intermediate representation, known as SlitheR, from the Solidity source codes. SlitheR provides in-depth information on inheritance relationships and control flows for manual and automated inspection.

We used \textit{MythX} to analyze source code.
MythX~\cite{MythXSmartcontractsecurityserviceforEthereum-2022-06-09} is a web-based platform that provides three tools for analyzing smart contracts: Mythril~\cite{mythril}, Harvey~\cite{wustholz2020harvey}, and Maru. It analyzes both the source code and compiled bytecode of smart contracts and combines the results of the three tools to improve accuracy. Mythril uses symbolic execution to explore all possible paths of a smart contract, while also applying taint analysis to more accurately identify known vulnerabilities. Harvey is a smart contract fuzzer that generates random inputs to test for unexpected behavior, and Maru performs static analysis on smart contracts. 

\section{Experiment Design}\label{Methodology-Experimental-Setup}

\subsection{Designing Prompts}\label{prompt-eng}
\textit{Prompt} is a function of the LLM that provides a set of instructions to the program to customize, enhance, or refine the program.

%In the context of LLMs, a set of instructions provided to the program, to customize, enhance or refine its capabilities is called a \textit{prompt}. 

Prompt plays a crucial role in shaping subsequent interactions and the output generated from an LLM by setting specific rules and guidelines for an LLM conversation. The prompt provides initial rules, sets the context for the conversation, and informs the LLM about what information is important, as well as what the desired output form and content should be. Prompt engineering is the process used to program LLMs using prompts.

The concept of prompt patterns is similar to that of software patterns in that it provides a codified approach to solving specific problems, but it focuses specifically on the context of output generation from LLMs such as ChatGPT. Prompt patterns provide a means to customize the output and interactions of LLMs, similar to how software patterns provide a codified approach to common software development challenges. By documenting and leveraging prompt patterns, users and teams can enforce constraints on the generated output, ensure that relevant information is included, and change the format of interaction with the LLM to better solve the problems that they are facing. 

Each prompt pattern follows a similar format to classic software patterns, but with slight modifications to match the context of output generation with LLMs. It includes a name and classification, the intent and context, the motivation, the structure and key ideas, an example implementation, and consequences. The prompt pattern name uniquely identifies the pattern and indicates the problem being addressed, while the classification includes categories of pattern types such as output customization, error identification, prompt improvement, interaction, and context control. The intent and context describe the problem that the prompt pattern solves, and the goals that it achieves, while the motivation provides the rationale for the problem and explains why solving it is important. The structure and key ideas section describe the fundamental contextual information that the prompt pattern provides to the LLM, while the example implementation demonstrates how the prompt pattern is worded in practice. Finally, the consequences section summarizes the pros and cons of applying the pattern and may guide how to adapt the prompt to different contexts.

For our experiments, we used a guided approach to generate our prompts. For a generation, we used the following formula for generating prompts:

\textbf{Step 1:} All the extracted codes were passed through a Python parser that would extract the code comments from each of the Solidity files along with names. These were then saved in an intermediate file with the file name.
\textbf{Step 2:} We would construct prompts based on the previously generated \textit{code template} and \textit{Step 1.}

\begin{itemize}
    \item \textbf{Google Palm2}:  For Palm2 the recipe we followed was: \textit{Can you write a secure program that does} \textbf{filename} on the \textbf{code template}.

    \item \textbf{ChatGPT}:  For GPT the recipe we followed was: \textit{Can you give example of a secure program that does} \textbf{filename} on the \textbf{code template}.
\end{itemize}
    
\begin{figure}[h]
    \centering
    \includegraphics[scale=0.35]{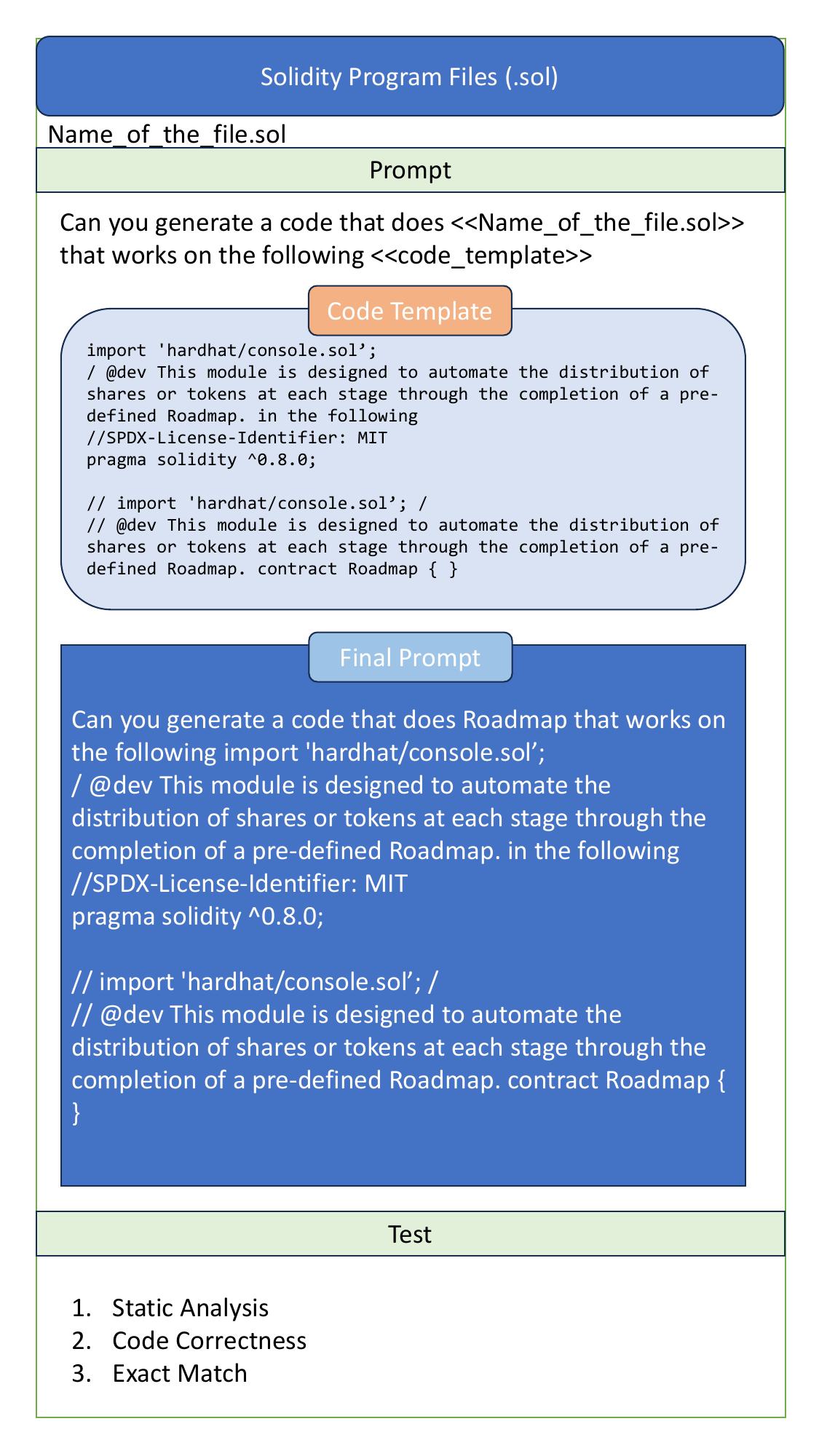}
    \caption{Prompt generation example.}
    \label{fig:prompteng}
\end{figure}
        
% # verbal description how prompt engineering for both AI tools
% # give some example, prompt used and compare these two AI tools
% # how code generated depends on the prompt
% # obervations example
% ----> ask chatgpt to generate code, say no vulernability, no overflow, 
% --------> does the code have overflow
% ----> ??
\subsection{Experimental Workflow}

\begin{figure}[ht]
\centering
\includegraphics[scale=0.5]{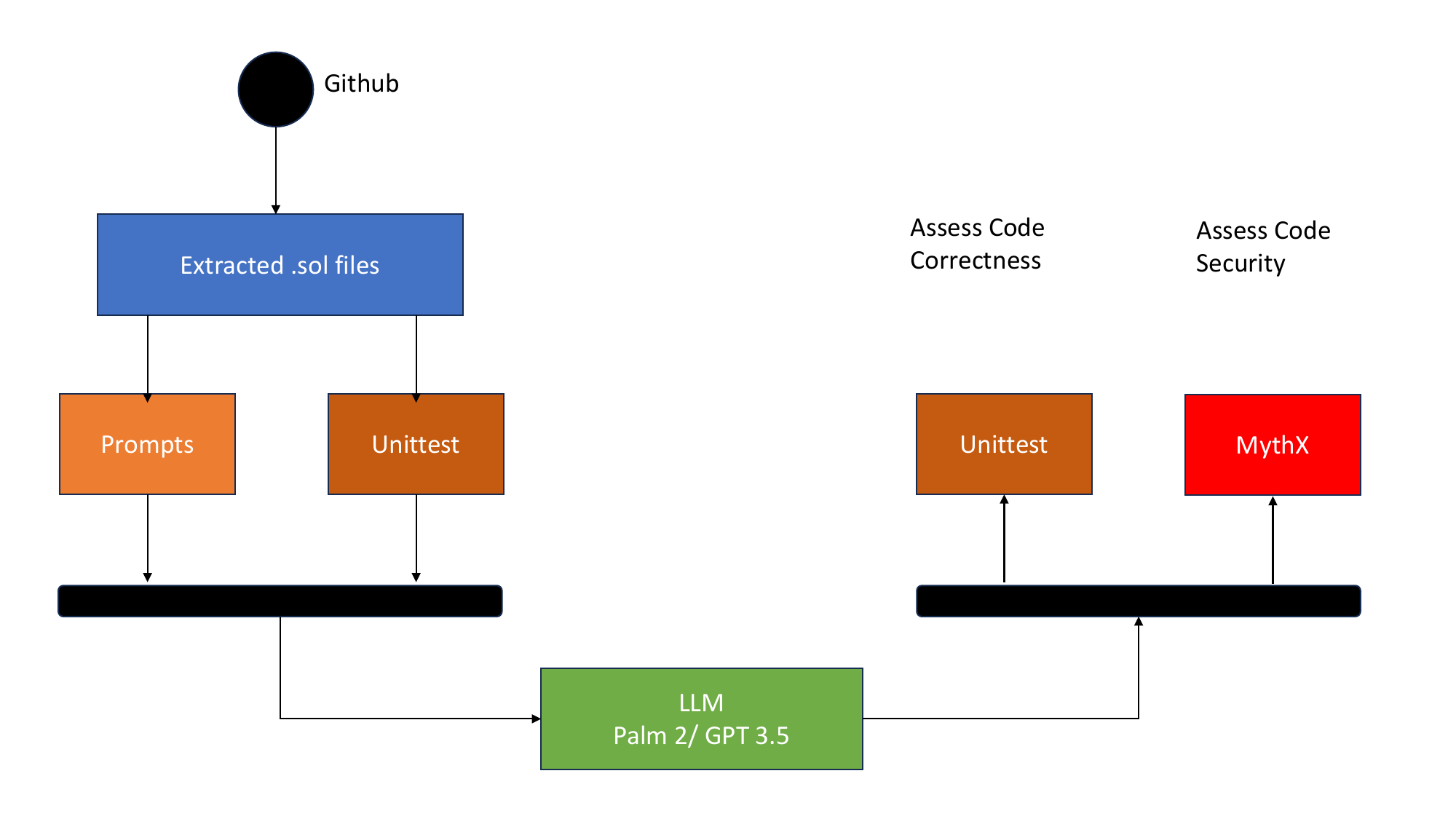}
\caption{A step-by-step illustration of the experiment's workflow.}
\label{experimental_workflow}
\end{figure}

In Figure \ref{experimental_workflow}, we provide a diagram of how the experiment was conducted. First, the dataset has been collected from Github and then pruned using a data cleaner script. This has been de-dpulicated to include only unique files. Then we use a combination of custom python script and solhint to extract comments from the .sol files. For each of the projects we evaluated, we also look at any unittest folder present in the repo. These are then used for generating prompt pairs using template. The unittest are kept in separate folders and tracked by Repo ID. The prompt generation has been detailed in Section \ref{prompt-eng}. Given the non-deterministic nature of the code generation tools, we inspect the prompts and then use Palm2 and Gpt 3.5 API to generate the codes with temperature 0. Once this is completed, we start the assessment phase by running unit tests on the codes to assess validity and correctness. After this, we utilize Mythx\cite{MythXSmartcontractsecurityserviceforEthereum-2022-06-09} to detect any potential bugs that might have been introduced. We detect the bugs and code smells separate, since the code smells are mostly different from bugs, they do not necessarily cause the code to be incorrect, but introduce some uneasiness in the code which can cause more problems in the future. We save the individual assessment results related to the problem for each step of the assessment phase.

\section{Evaluation}

\subsection{Code Metrics}\label{Methodology-Our-Metrics}
We have assessed our results based on several metrics related to the quality of the generated Solidity code, including code validity, code correctness, code security, code reliability, and code maintainability.

To evaluate code validity, we use a binary metric with two possible values, 0 and 1, indicating whether the generated code is valid or not. We assess code validity by checking how well the code adheres to the syntax rules of Solidity and whether any errors occur during runtime. We use the solc compiler to check for code validity since our dataset is constructed for the Solidity programming language. After initiating the program, we record any errors that could be raised during runtime.

For code correctness, we measure the extent to which the generated code performs as intended. We use problem-specific unit tests %cherry-picked 
from our GitHub Dataset to assess code correctness. On average, each problem comes with 7 unit tests, and we calculate code correctness as the number of passed unit tests divided by the total number of unit tests for a specific problem. We believe that this is the most convenient way to assess code correctness, given their abundance.

We also evaluate the average code correctness, which is measured as the sum of all code correctness scores divided by the problem count.

In addition to the metrics mentioned above, we also evaluated the following:

\begin{itemize}
    \item \textit{Code security}: We used MythX to assess the security of the generated code. MythX is a static analysis tool that can identify security vulnerabilities in code.
    \item \textit{Code reliability}: We used the number of bugs detected by MythX to assess the reliability of the generated code. Bugs are errors in code that can cause the code to malfunction. Code smells \cite{yeticstiren2023evaluating} are patterns in code that can indicate potential problems, but they do not necessarily cause the code to be incorrect.
    \item \textit{Code maintainability}: We used the number of comments in the generated code to assess the maintainability of the code. Comments are used to explain the code, which can make it easier to understand and maintain.
\end{itemize}

We show our calculation methods for Code Correctness and Average Code Correctness as defined by \cite{yeticstiren2023evaluating} below. $CCS$ stands for Code Correctness Score, and $CCS_{i}$ is the Code Correctness Score for the $i\textsuperscript{th}$ problem. The range of $i$ is 0 to n.
            $$
            Code\:Correctness = \frac{\sum_{i=0}^{n}{CCS_{i}\:[CCS_{i}=1]}}{n}
            $$
            \\
            $$
            Average\:Code\:Correctness = \frac{\sum_{i=0}^{n}{CCS_{i}}}{n} 
            $$
            \\

Here n=86 for the final dataset for the codes that were compilable for both ChatGPT and Palm2. However, we can generalize the formula to other scenarios.

In the final stages of our evaluation, we use MythX to assess the security, reliability, and maintainability of the generated code. We define a vulnerability as a weakness in the code that could be exploited by an attacker to gain unauthorized access to the system. Vulnerabilities can introduce potential risks, whether the code is deployed as standalone software or as part of a larger software system.

\section{Results}\label{Results}
        \subsection{Code Validity (RQ1.1)}\label{Results-Validity}

The use of solc compiler made it easy for us to evaluate code validity just by trying to execute the generated codes. We were able to log all successful and unsuccessful generations and analyze them. Compile errors were also logged for guided code generation for our future work. 

\begin{figure}[ht]
\centering
\includegraphics[scale=0.06]{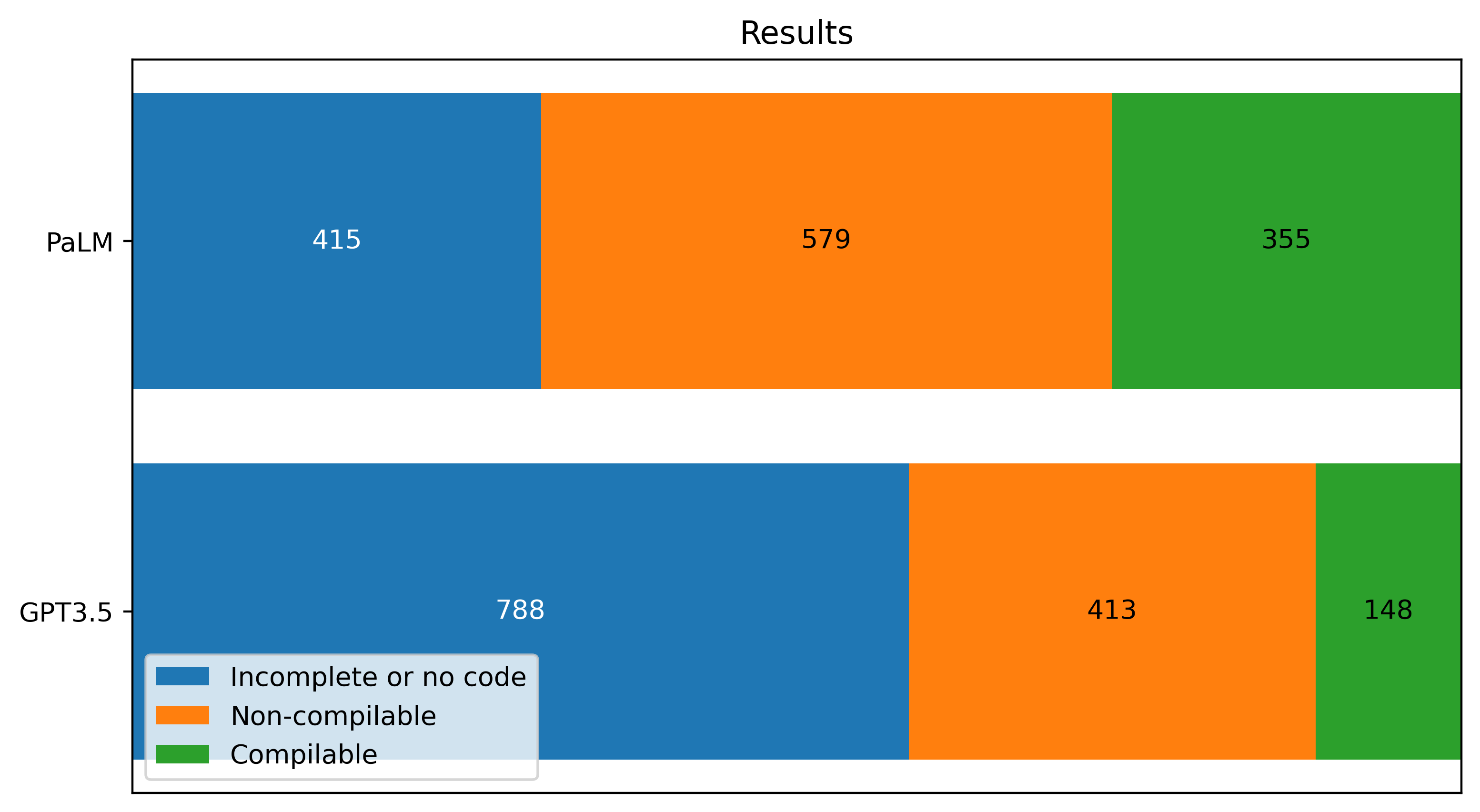}
\caption{Code compile statistics.}
\label{fig:codecompile}
\end{figure}

\subsection{Code Correctness (RQ1.2)}\label{Results-Code-Correctness}
We used the number of passed unit tests divided by all unit tests to calculate the success percentage of the code for each problem, as defined in Section \ref{Methodology-Our-Metrics}. 

We also measured the average code correctness score by dividing the summation of all code correctness scores by the number of all problems, as defined in Section \ref{Methodology-Our-Metrics}. 

% \begin{minipage}{\linewidth}
%     \begin{tikzpicture}
%         \pie[rotate=90, color={green!60, yellow!60, red!60}]{
%             46.3/Correct,
%             23.2/Partially Correct,
%             30.5/Incorrect}
%     \end{tikzpicture}
% \end{minipage}

\begin{figure}[ht]
\centering
\includegraphics[width=\linewidth]{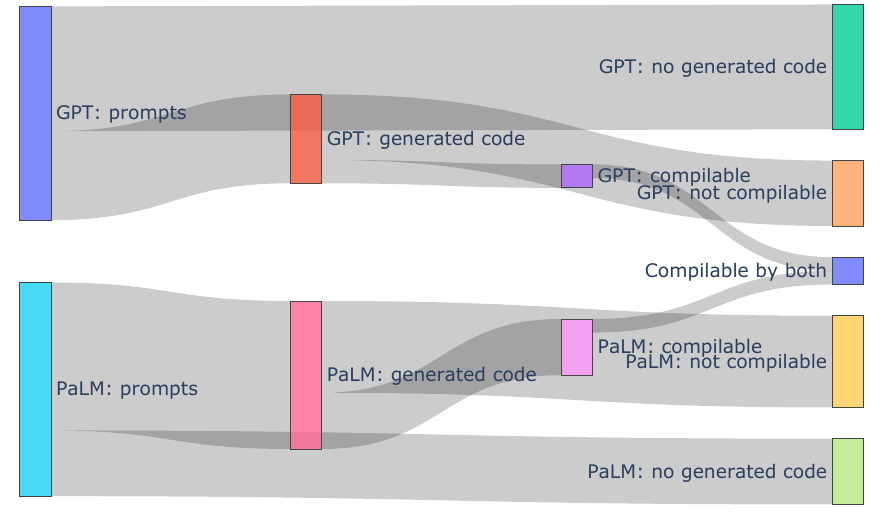}
\caption{Division of different generated codes based on their correctness.}
\label{fig:codeshanky}
\end{figure}

\subsection{Code Generation Improvement (RQ2)}\label{Results-Using-Dummy-Function-Names-(RQ3)}

In this part, as explained in Section \ref{Methodology-Experimental-Setup}, we prompted ChatGPT and Palm2 to generate code for the same problems, this time with placeholder function names instead of meaningful, and informative function names, and also without any comment information, without code template. We replaced the function names with `waldo'.

Our code validity score increased to 93.9\% for Palm2 and decreased to 89.6\% for ChatGPT.

For code correctness, if we compare the results of the two experiments for ChatGPT, the rate of correctly generated code dropped by 4.2\%. The incorrectly generated code percentage increased from 0.6\%, and the partially correctly generated code percentage increased from 4.6\%. For Palm2, the rate of correctly generated code dropped by 3.6\%. The incorrectly generated code percentage increased from 0.6\%, and the partially correctly generated code percentage increased from 3\%.

\subsection{Code Security(RQ3)}\label{Discussion-Code-Security-Reliability-Maintainability}
\begin{figure}
    \centering
    \includegraphics[scale=0.06]{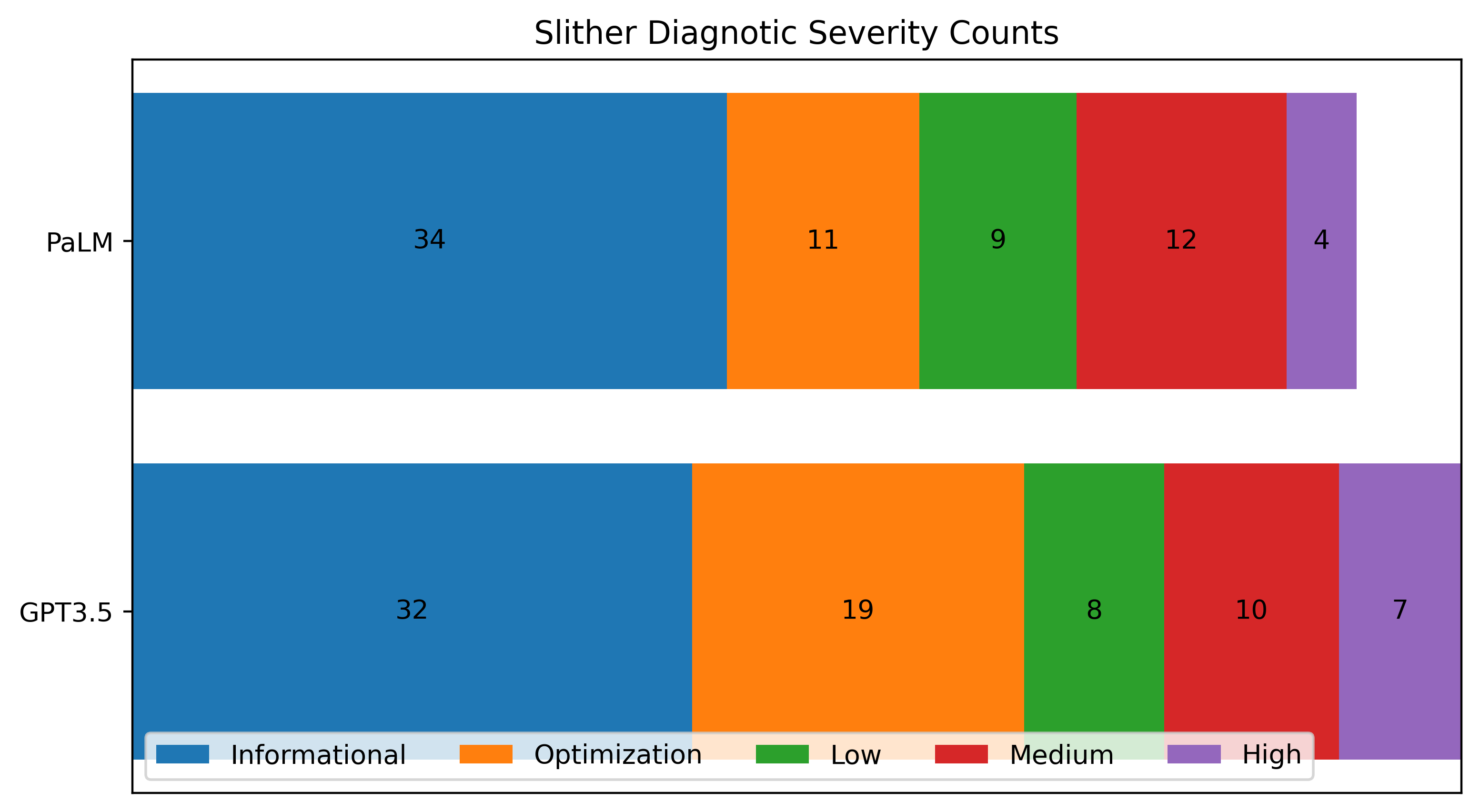}
    \caption{Security incidents reported by Slither.}
    \label{fig:slither-bug}
\end{figure}

\begin{figure}
    \centering
    \includegraphics[scale=0.06]{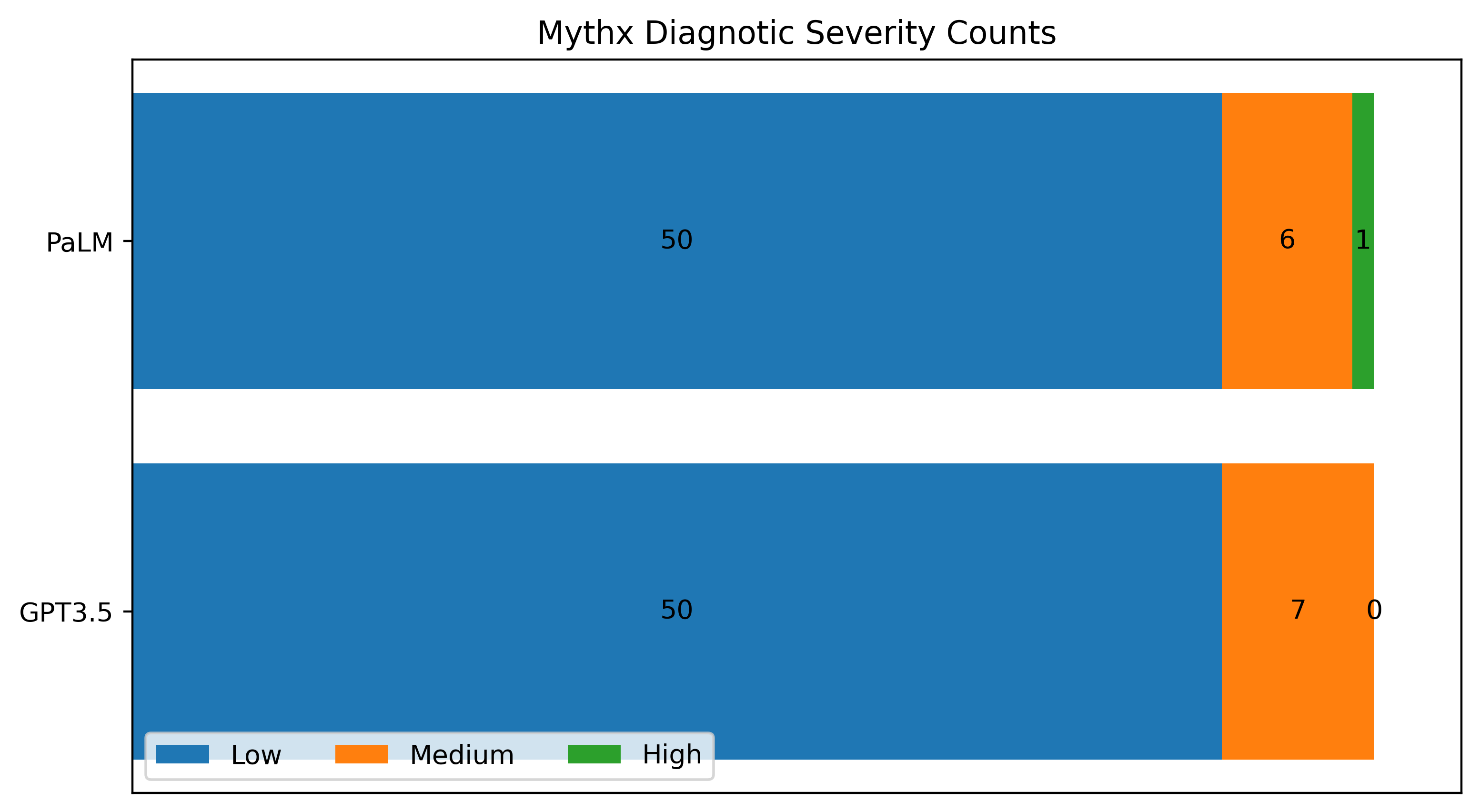}
    \caption{Security incidents reported by MythX.}
    \label{fig:mythx_bug}
\end{figure}

As explained in Section~\ref{sec-dis}, the generated codes had different bugs that got reported by both Slither and Mythx.
\figurename~\ref{fig:slither-bug} and \figurename~\ref{fig:mythx_bug} show the distribution of different severity bugs among the generated code between two LLMs.

A detailed distribution of the security incidents has been described in \tablename~\ref{slither-table} and \tablename~\ref{mythx-table}.

\medskip
        \fbox{\begin{minipage}{23em}
           Surprisingly it seems Slither catches more High Severity bugs in code produced by ChatGPT than MythX. However, MythX shows a marginal increase in severe bugs produced in Palm2-generated Code.
        \end{minipage}}     
\begin{table}[H]
\centering
\caption{\label{slither-table}Bugs found by Slither}
\resizebox{\columnwidth}{!}{%
\footnotesize
\begin{tabular}{llrr}
\toprule
&  & GPT3.5 & PaLM2 \\
Severity & Lint &  &  \\
\midrule
\multirow[t]{7}{*}{Informational} & dead-code & 24 & 27 \\
 & similar-names & 0 & 2 \\
 & assembly & 4 & 0 \\
 & reentrancy-unlimited-gas & 1 & 0 \\
 & low-level-calls & 1 & 1 \\
 & too-many-digits & 1 & 1 \\
 & unused-state & 1 & 3 \\
\cline{1-4}
\multirow[t]{2}{*}{Optimization} & constable-states & 7 & 7 \\
 & immutable-states & 12 & 4 \\
\cline{1-4}
\multirow[t]{6}{*}{Low} & reentrancy-benign & 1 & 0 \\
 & calls-loop & 1 & 1 \\
 & missing-zero-check & 2 & 1 \\
 & reentrancy-events & 0 & 1 \\
 & timestamp & 3 & 3 \\
 & shadowing-local & 1 & 3 \\
\cline{1-4}
\multirow[t]{4}{*}{Medium} & erc20-interface & 0 & 5 \\
 & incorrect-equality & 5 & 4 \\
 & locked-ether & 2 & 3 \\
 & tautology & 3 & 0 \\
\cline{1-4}
\multirow[t]{2}{*}{High} & uninitialized-state & 4 & 0 \\
 & weak-prng & 3 & 4 \\
\cline{1-4}
\bottomrule
\end{tabular}
}

\end{table}

\begin{table}
\centering
\caption{\label{mythx-table}Bugs found by MythX}
\resizebox{\columnwidth}{!}{%
%\scriptsize
%\footnotesize
\begin{tabular}{llrr}
\toprule
 &  & GPT3.5 & PaLM2 \\
Severity & Lint &  &  \\
\midrule
\multirow[t]{7}{*}{Low} & Assert Violation & 25 & 28 \\
 & State Variable Default Visibility & 5 & 15 \\
 & Weak Sources of Randomness from Chain Attributes & 13 & 3 \\
 & Authorization through tx.origin & 5 & 3 \\
 & Timestamp Dependence & 1 & 2 \\
 & Reentrancy & 1 & 0 \\
 & Requirement Violation & 0 & 1 \\
\cline{1-4}
\multirow[t]{2}{*}{Medium} & Incorrect Constructor Name & 6 & 5 \\
 & DoS with Failed Call & 1 & 1 \\
\cline{1-4}
High & Integer Overflow and Underflow & 2 & 3 \\
\cline{1-4}
\bottomrule
\end{tabular}
}
\end{table}

\section{Security Analysis}\label{sec-dis}

\subsection{Security Analysis}
In this section, we demonstrate several code segments that were generated and have one or more specific security issues.
We highlight some of the bugs we detected in these codes in the following sections. For all of these checks we used the tools from Section~\ref{tool}.

\para{Timestamp dependency.} 
The concept of timestamp dependency refers to the use of the current timestamp of the Ethereum block by smart contracts to process time-dependent events. However, since the timestamp can be manipulated to some extent by miners, malicious miners may take advantage of this vulnerability to alter the timestamp for their own benefit. This bug was used in the governmental Ponzi scheme attack. As a result, developers should not rely solely on the precision of the block's timestamp. %An example of a code snippet that represents this bug is shown in Figure 4, where block. timestamp is used to return the block's timestamp.

\begin{lstlisting}[language=Solidity,basicstyle=\footnotesize,caption=Timestamp dependency,label={lst:timestamp}]
pragma solidity ^0.8.0;
contract TimestampDependency {
    uint256 public lastTime;
    uint256 public counter;
    function increment() public {
        require(block.timestamp > lastTime + 1 minutes, "Must wait at least 1 minute");
        lastTime = block.timestamp;
        counter++;
    }
}
\end{lstlisting}

In Listing~\ref{lst:timestamp}, the increment function increases a counter variable, but it only allows one increment per minute. The function uses the \texttt{block.timestamp} value to check when the last increment occurred.

However, this approach is vulnerable to manipulation by miners. A malicious miner could modify the timestamp of the block to allow multiple increments within one minute, breaking the intended behavior of the contract.

\para{Re-entrancy.}
In Solidity, contracts can have external calls that can be accessed by attackers to execute a function within the same contract multiple times, leading to unexpected operations within the contract. The DAO attack in 2016 was a notable instance that exploited this vulnerability.

\begin{lstlisting}[language=Solidity,basicstyle=\footnotesize,caption=Re-entrancy,label={lst:bank}]
pragma solidity ^0.8.0;
contract Bank {
    mapping (address => uint) private balances;
    function deposit() public payable {
        balances[msg.sender] += msg.value;
    }
    function withdraw(uint amount) public {
        require(balances[msg.sender] >= amount, "Insufficient balance");
        (bool success,) = msg.sender.call{value: amount}("");
        require(success, "Transaction failed");
        balances[msg.sender] -= amount;
    }
    function getBalance() public view returns (uint) {
        return balances[msg.sender];}}
\end{lstlisting}

In this code Listing~\ref{lst:bank}, the withdraw function allows a user to withdraw a certain amount of Ether from their account balance. However, the function contains a re-entrancy bug because it performs an external call to msg.sender before updating the balance, which allows an attacker to call the withdraw function repeatedly before the balance is updated, potentially causing them to withdraw more Ether than they have in their account.

%An attacker can exploit this bug by deploying a malicious contract that calls the withdraw function repeatedly before the balance is updated, effectively draining the bank's balance. The DAO attack, which resulted in the loss of millions of dollars worth of Ether, was caused by a similar re-entrancy bug in a smart contract.

\para{Unchecked send.} 
The vulnerability known as "Unauthorized Ether transfer" occurs when certain functions in a contract are visible to the public and allow non-zero sends, allowing external users to transfer Ether from the vulnerable contract even if they lack the necessary credentials. As a result, unauthorized individuals can call these functions and transfer Ether without proper authorization.

\textit{We have noticed this multiple times in ChatGPT-generated code, especially for Dutch auction. %Which we manually changed so the newer compilation tools could run. 
}

\iffalse
\begin{lstlisting}[language=Solidity,basicstyle=\footnotesize,caption=Unchecked Send]
pragma solidity ^0.8.0;
contract UncheckedSend {
    mapping(address => uint) balances;  
    function deposit() public payable {
        balances[msg.sender] += msg.value;    }    
    function withdraw(uint amount) public {
        require(balances[msg.sender] >= amount, "Insufficient balance");
        (bool success,) =msg.sender.call{value: amount}("");
        if (!success) {
            // Handle failed send operation
        }
        balances[msg.sender] -= amount;}}
\end{lstlisting}
\fi

\begin{lstlisting}[language=Solidity,basicstyle=\footnotesize,caption=Unchecked Send]
pragma solidity ^0.8.0;
contract UncheckedSend {
    mapping(address => uint) balances;  
    function deposit() public payable {
        balances[msg.sender] += msg.value;    }    
    function withdraw(uint amount) public {
        require(balances[msg.sender] >= amount, "Insufficient balance");
        msg.sender.call{value: amount}("");
        balances[msg.sender] -= amount; }}
\end{lstlisting}

In the withdraw function, the contract transfers the requested amount of Ether to the caller using the call function. However, this send operation is unchecked, which means that if the call function fails for some reason, such as the receiver not having enough gas to execute its fallback function, the transaction will not be reverted and the Ether will be lost. This can result in unauthorized Ether transfers if the function is called by an attacker with insufficient balance or other invalid parameters.

\fbox{\begin{minipage}{23em}
All of these have been reported in the LLMs generated code among a plethora of other bugs as described in Section \ref{Discussion-Code-Security-Reliability-Maintainability} in detail.
\end{minipage}}

% \subsubsection{Designing the tests}

% # description of the tools
% # results from the tools
% # table? warning, error message from the tools 

\subsection{Performance and Smart Contract Specific Analysis}
% # high level, your assessment how well written the code is
% have comment? no comment? ...
% # number of lines (code)
% # gas fees
In addition to correctness and security assessment, we have evaluated GPT-generated contracts using certain smart contract-specific metrics like contract size and gas fee cost. 

We measured the bytecode size of the smart contracts generated by the GPT models. Compared with the handwritten contracts, the contracts generated by the GPT models may have smaller size. For instance, a manually written auction contract has bytecode size of 4.747KB. The ChatGPT generated version has size of 2.286KB. %The handwritten contracts often contain more code like modifiers, attribute access functions, requirement checks, etc.  

Since the GPT models were trained with the source code dataset publicly available, when asked to generate a contract that is likely contained in the training dataset, the GPT models may generate contract code similar to the original code in the training dataset (with different variable names and/or function names). For instance, we have tested the GPT models to generate contract for Black-Scholes. The contract size by ChatGPT is 2.371KB, which is very close to the publicly available version that has 2.399KB bytecode size. 

To evaluate the run-time performance of the GPT model-generated smart contracts, we measured gas cost for running smart contract functions. For certain tests, running AI-generated smart contract functions cost less gas than the manually written functions used for comparison. %Handwritten functions may cost more gas because these functions typically contain additional code for requirement checking, modifier, or extra code for exception handling. 
To collect the gas fee for each function, we applied a Solidity contract fuzzer~\cite{9240669}. The fuzzer, implemented on top of PyEVM, can analyze the source code and generate test cases for each function in a contract. At the same time, it measured gas fee cost, each time a function was called. 

Table ~\ref{gas} shows the gas cost results for two auction contracts, one created by ChatGPT and the other one manually written. The table lists min and max gas fees for four functions of each contract. Handwritten functions used more gas on average, due to the presence of additional code and abort logic. ChatGPT-generated functions did not contain this additional code, resulting in lower gas costs.

%\vspace{-9pt}
\begin{table}
    \caption{Gas cost comparison for auction contracts.}
    \label{gas}
    \scriptsize
    \centering
    \begin{adjustbox}{width=\columnwidth,center}
    \begin{tabular}{|p{1.3cm}|p{1.3cm}|p{1.3cm}|p{1.3cm}|p{1.3cm}|}
    \hline
    & \multicolumn{4}{|c|}{Contracts} \\ \cline{2-5}
    &\multicolumn{2}{|c|}{ChatGPT}& \multicolumn{2}{|c|}{Manual} \\ \hline
    & Min & Max & Min & Max \\ \hline
    bid & 2678 & 43075 & 0 & 66135 \\ \hline
    highestBid	 & 526 & 526 & 0 & 816 \\ \hline
    highestBidder	 & 486 & 486 & 0 & 596 \\ \hline
    endPrice & 504 & 504  & 0 & 658 \\ \hline  
    \end{tabular}
    \end{adjustbox}

 \end{table}
%\vspace{-20pt}

We observed that GPT models produce contract code that is typically smaller and less gas-intensive than code written by human developers. %This was because the handwritten code typically contained modifiers, additional code for requirement checking, more attribute access functions, etc. 
% In case the GPT models were asked to generate a contract similar to what was used as training data, the generated code might resemble the contracts contained in the training data with completely renamed variables and/or function names. In this case, the size and gas fee cost might be similar to the training dataset.

\section{Threats to Validity of Research Design}
While our approach shares some similarities with previous studies, this work is more thorough than any other approach we are aware of for Solidity code. It is important to note, however, that current experiments have some limitations.
%Our approach is similar to previous studies, but it is more thorough than any other approach we are aware of for Solidity code. It is important to note, however, that this experiment has some limitations.

\para{One-shot code generation.} 
When we used code generators to generate code, we provided them with the function names, parameters, and a docstring explaining the function. We also provided them with a few test cases. However, we did not provide them with any additional code snippets that would clarify our intent for the specific problem. In most cases, the success rate of the code generators could be improved if we provided them with hints in the form of code snippets.
          
\para{Reproduction of the generations.} 
When we ran our experiments, we noticed that the code generators were not deterministic. This means that they would generate different outputs for the same input in different trials. To avoid this, we generated code for each problem in one iteration, saved the code, and then evaluated the saved code. However, even with this precaution, our results may not be fully reproducible because the underlying LLMs of the code generators are being retrained over time.

\para{Metrics.} 
As outlined in Section~\ref{Methodology-Our-Metrics}, we assessed the generated code utilizing the Code Validity, Correctness, and Security metrics. Nevertheless, we acknowledge the possibility of incorporating supplementary metrics, including Readability, Cyclomatic Complexity, and Reusability, to comprehensively evaluate the generated code.

\section{Related Works}
Since the time when the GPT models were made open to the public, the models have attracted significant interest from the research community. Researchers have evaluated these models for generating source code in particular from a security vulnerability perspective ~\cite{https://doi.org/10.48550/arxiv.2108.09293,https://doi.org/10.48550/arxiv.2302.04012}. 
\cite{asleep} evaluated the security of the programs generated by GitHub Copilot. They found that 40\% of the generated programs were vulnerable. These results differ from our results, which may be due to the characteristics of our dataset.

Vaithilingam et al \cite{copilot-user-exp} conducted a user study to assess the effects of GitHub Copilot.
\cite{nguyen} evaluated GitHub Copilot using 33 LeetCode questions and four programming languages: Python, Java, JavaScript, and C. This also fundamentally differs from this work based on the programming language.
\cite{mastropaolo2023robustness} conducted an empirical study to investigate the effect of semantic-preserving changes in natural language on the generated code function of GitHub Copilot. They provided GitHub Copilot with 892 non-trivial Java method descriptions. In addition, researchers also explored LLMs for test code generation for languages such as Java ~\cite{nie2023learning}.

However, most of the existing work has been focused on languages such as Python, JavaScript, etc.  There was a gap in exploring the GPT models for smart contract generation in a comprehensive manner. This preliminary study attempted to fill this gap and highlighted some initial findings. Our research scratched the surface of this emerging research area that contains several potential directions for further investigation as pointed out in the analysis section. 

\section{Conclusion}

This work investigated two state-of-the-art code generation LLMs, namely OpenAI ChatGPT and Google Palm2 for generating smart contract code in Solidity. We evaluated the quality of the generated code in terms of correctness, validity, and security.
% To do that we also had to collect and introduce our own Solidity code dataset collected from Github.
We propose and demonstrate a validation framework for evaluating code quality for smart contract code, as well as a pipeline using which code verification can be mostly automated. 

In addition, we proposed a code template-based code generation approach using prompts that help us achieve better compilable code than prompts without context. 

Based on the findings, for now, Palm2 produces consistently more compilable code than ChatGPT. However, they are matched when it comes to introducing bugs in the generated code. This poses a problem for recommending any of these tools to actual programmers.

Future work will consist of improving upon the verification platform so that it can handle a much bigger dataset instead of a curated popular dataset. Furthermore, we aim to extend our template-based code generation to provide a more robust, universal, and safe way to generate smart contracts using LLMs.

\section*{Author Contributions}
\insertcreditsstatement

\bibliographystyle{IEEEtran}
\bibliography{ref}
%
% \begin{thebibliography}
% \end{thebibliography}
\end{document}